\documentclass[prl,twocolumn]{revtex4}
\usepackage{graphicx}

\begin{document}

\title{First-Principles Computations for KTN}
\author{S. Prosandeev$^{1,2}$, E. Cockayne$^1$, B. Burton$^1$ and
A. Turik$^2$}  \affiliation{$^1$National Institute of Standards
and Technology, MD; $^2$Rostov State University} \maketitle

Solid solutions of KTaO$_{3}$ and KNbO$_{3}$ (KTN) exhibit unusual
dielectric properties at ferroelectric phase transitions, which
are usually associated with a Nb instability \cite{Kleemann} that
is often explained by the large Nb dynamical charge in oxides.
Using non-self-consistent VASP computations \cite{Washington}, we
showed earlier that the Nb dynamical charge in layerd KTN strongly
depends on the local environment of Nb. In the present study, we
have performed self-consistent first-principles computations for
fixed composition KTa$_{7 / 8}$Nb$_{1 / 8}$O$_{3}$ in different
configurations shown in Fig. 1. We computed the equilibrium ionic
positions, Born charges, force constants, vibration frequencies
and modes.

In agreement with our previous results, we found that the Nb
dynamical charge is sensitive to the nearest neighbors (Table 1).
The Nb charge in directions where there are only Nb ions is about
9.9e. In the direction where there are Ta ions, the Nb dynamical
charge is about 8.3e. In contrast, the Ta charge increases in
directions where there are Nb ions. The comparison of the energies
(Table 2) shows that there is tendency to Nb ordering in planes if
one does not take into account external stress necessary for
obtaining experimental lattice constant in KTaO$_{3}$. When we
applied this stress, we obtained that the energies of the
considered structures are close to each other. Note that this
situation is very different from that known for PST, PMN etc.: in
those cases, 1:1 or 1:2 order appears as a result of electrostatic
interactions due to different charges of Nb and Mg or Sc and Ta
\cite{Burton}. In KTN, the valence of Nb and Ta is the same.

\begin{figure}
\resizebox{0.45\textwidth}{!} {\includegraphics{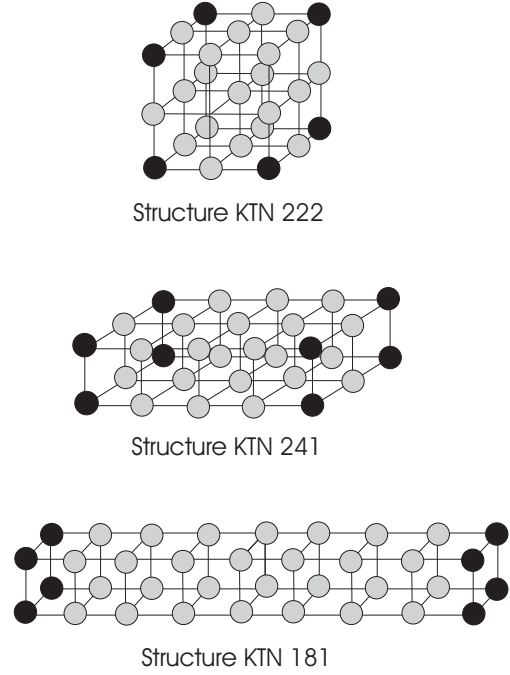}}
\caption{The geometry of the structures considered.} \label{x}
\end{figure}

\begin{table}[htbp]
\caption{Ta and Nb dynamical charges in KTN 181. The numbers
correspond to the Ta ions from the left to the right (Fig. 1).}
\begin{ruledtabular}
\begin{tabular}{c|ccc|c|ccc}
 & x& y& z & & x& y& z
\\ \hline 1Ta& 8.758& 8.938& 8.758 & 2Ta& 8.755& 9.009&
8.755
\\ 3Ta& 8.753& 9.017& 8.753 &  4Ta& 8.752& 9.017&
8.752 \\ 5Ta& 8.753& 9.017& 8.753 &  6Ta& 8.755& 9.009& 8.755
\\ 7Ta& 8.758& 8.938& 8.758 &  Nb& 9.898& 8.259& 9.898
\end{tabular}
\end{ruledtabular}
\label{tab1}
\end{table}

\begin{table}[htbp]
\caption{Total energies (with respect to the highest value
obtained) for the computed supercells [eV]. ``equilibr.'' is the
self-consistent energy obtained at zero stress. ``Vegard'' is the
energy obtained at the volume which is the average between
KTaO$_{3}$ and KNbO$_{3}$. ``KT-stress'' is the energy obtained at
the hydrostatic stress giving experimental lattice constant for
KTaO$_{3}$.}
\begin{ruledtabular}
\begin{tabular}{c|ccc}
structure& KTN222& KTN181& KTN241 \\ \hline equilibr.& -0.08082&
-0.38057& -0.08071
\\ Vegard& -0.06353& -0.32842& -0.06334 \\
KT-stress& -0.00250& -0.00000&
\end{tabular}
\end{ruledtabular}
\label{tab2}
\end{table}

\begin{table}[htbp]
\caption{Vibration frequencies obtained for TO1 soft modes [cm$^{
- 1}$] at the external stress corresponding to an experimental
lattice constant in KTaO$_{3}$.} \begin{ruledtabular}
\begin{tabular}{cc|cc}
\multicolumn{2}{c}{KTN 222} & \multicolumn{2}{c}{KTN 181}
\\ \hline $\nu$ & & $\nu$ &  \\ \hline
 44,\,\,54 & x & 8i,\,\,30 & z \\
 44,\,\,54 & y & 8i,\,\,30, & x \\
 44,\,\,54 & z & 35,\,\,41 & y
\end{tabular}
\end{ruledtabular}
\end{table}

We obtained that the average Nb-O distance is larger than the
average Ta-O distance. It implies that the Nb ions cause tensile
distortion in KTN. Nevertheless, the lowest soft TO modes are
mostly due to Nb (Table 3).

We also found that the lattice instability in the KTN 181
configuration appears in the directions of the Nb planes. That is
consistent with other first-principles computations
\cite{Sepliarsky}. The instabilities can be explained by the
long-range dipole-dipole interaction supplemented by the covalent
(pseudo vibronic) effect: the displacement of one of the Nb ions
along the Nb-O-Nb bond triggers appearing an additional force on
nearest Nb ions in the same chain due to the change of electronic
structure that is produced by the shift of the Nb ion
\cite{Harrison}: $\varepsilon _{cor} (R_1 ,R_2 ) =
\sum\limits_{\tau,\tau'} {
 \left[W(R_1 )_{\tau,\tau'} W(R_2 )_{\tau',\tau}\right]
  f^\tau \left( 1 - f^{\tau'}\right)/
  \left[\varepsilon^\tau- \varepsilon ^{\tau '}\right]}$~
where $W(R) = r\left( {\partial H / \partial R} \right)$ is the
linear vibronic operator, $r$ being the ionic displacement and $H$
being the electron Hamiltonian, $\varepsilon ^\tau ,\,f^\tau$~ are
electron energy and occupation number, respectively, $\tau $
denotes the electronic state.

By using a tight-binding approach \cite{Nerodo,Harrison} we have
computed the correlation energy for the ionic displacements in
KNbO$_{3}$, in $(z^2/R_{\mathrm{Nb} - \mathrm{O}}^2
)\,\mathrm{eV}$ units, where $z$~ being the ionic charge and $R=
R_{Nb - O}$ is the Nb-O bond length. When the nearest Nb ions are
shifted, the correlation energy is --10.3 and --0.9 for the
displacements along and perpendicular to the Nb-O-Nb line,
respectively. If Nb and nearest O are shifted then the correlation
energy is 25.9 and 8.6 for the displacemnts along and
perpendicular to the Nb-O bond, respectively. Hence, the vibronic
effect facilitates the displacement of the Nb ions in the chains
in the same direction while the Nb and O ions are easily shifted
in opposite directions. This result is in qualitative agreement
with straightforward computations of the interatomic force
constants (see \cite{Tinte} and references therein), which showed
that the M-O force constant along the \ldots -M-O-M-O-\ldots
chains proves to be practically negligible while the M-M force
constant appears to be surprisingly strong.

Hence, the local polarization in KTN appears in the Nb-reached
regions. The polarization direction follows the field and the free
energy can be written as: $F=\alpha (P-P_0)^2 + A (P-P_0)^3 + ...
- (E+E_0)P$~ where $E_0$~ is random and $P$~ is polarization
\emph{magnitude} ($P>P_0>0$~ where $P_0$ is the lowest possible,
remnant, magnitude of polarization). An unusual field dependence
of susceptibility follows from this:
$1/\chi(E)=2\alpha+3AE/\alpha+...,\,E \gg E_0$. At smaller fields,
susceptibility is constant or decreases because of averaging over
$E_0$~ \cite{hydro}.

The strongly angle dependent correlation of the displacements can
explain the diffuse X-ray scattering in KNbO$_{3}$~ \cite{Comes}
without the assumption that the Nb ions have \textit{deep}
off-center wells. At high temperatures (where the soft mode is
stabilized), acoustic excitations of the (Nb or O) chains can be
intensive due to nearly flat transverse acoustic phonon energy
surface in (001) planes \cite{Dorner}. The peaks in the XANES data
\cite{Ravel} can be understood if the large chain-like
fluctuations are considered together with the strong dependence of
the X-Ray absorption matrix element on the ionic displacements
\cite{Vedrinskii}.

The dynamical nature of the cooperative displacements
\cite{Takahasi} is supported by the fact that, in the high
temperature cubic phase, there is little dielectric dispersion
\cite{Turik}. In the ferroelectric phases, the situation is more
complicated. For instance, in the tetragonal phase, the average
polarization is directed along the [001] direction. However, the
direction of a local Nb displacement is [111] due to
anharmonicity. The correlation of the displacements can result in
the appearance of finite size chains directed in the [001]
directions, while the metal ions are displaced in [111] directions
(common within a finite-size chain but random in different chains
\cite{Takahasi} ). One can introduce population numbers for such
chains corresponding to different directions of the displacements.
Note that these (quasistatic) chains can exist only because of i)
anharmonicity of the potential relief, where the direction of the
(local) displacement does not coincide with the direction of the
polarization, ii) strong correlation of the displacements, iii)
existence of a uniform macroscopic field. In the cubic phase, the
last condition fails and the dispersion connected with the chains
disappears. In PbTiO$_3$, the low-temperature phase is tetragonal:
the ions are shifted in the [001] direction and the first
condition fails; the dispersion is also absent in accordance with
experiment \cite{Turik}. In the ferroelectric phases of BaTiO$_3$~
and KNbO$_3$, all three conditions are satisfied and the
dispersion exists \cite{Turik}.


\begin{thebibliography}{999}



\bibitem{Kleemann}  W. Kleemann, F.J. Schafer and D. Rytz, Phys. Rev. Lett. 54,
2038 (1985).

\bibitem{Washington}S. Prosandeev, E. Cockayne, and B. Burton,
AIP Conference Proceedings \textbf{626}, 64 (2002);
cond-mat/0208316.

\bibitem{Burton}B.P. Burton and E. Cockayne Phys. Rev. B \textbf{60}, R12542
(1999).
\bibitem{Sepliarsky} M. Sepliarsky, S.R. Phillpot, D. Wolf, M.G. Stachiotti and
R.L. Migoni, J. Appl. Phys. \textbf{90}, 4509 (2001).
\bibitem{Harrison} W.A. Harrison, Electronic structure and
properties of solids, Ed. W.H. Freeman \& Co., San Francisco 1980.

\bibitem{Nerodo} S.A. Prosandeev and A.A. Nerodo, Izv. ANSSR, ser. fiz.
\textbf{57}, 73 (1993).

\bibitem{Tinte}S. Tinte, M.G. Stachiotti, M. Sepliarsky, R.L. Migoni and C.O.
Rodriguez, J. Phys.: Condens. Matter \textbf{11}, 9679 (1999).

\bibitem{hydro}S.A. Prosandeev, Phys. Sol. St. \textbf{45} 1774
(2003).

\bibitem{Comes}R. Comes, M. Lambert and A. Guinier. Acta Cryst.
A\textbf{26}, 244 (1970).

\bibitem{Dorner}B. Dorner and R. Comes, in Dynamics of Solids and Liquids by Neutron
Scattering, ed. S.W. Lovesey and T. Springer, Springer-Verlag,
Berlin, 1977.

\bibitem{Ravel}B. Ravel, E.A. Stern, Physica B, \textbf{208/209}, 316
(1995).

\bibitem{Vedrinskii}R.V. Vedrinskii, V.L. Kraizman, A.A. Novakovich, Ph.V. Demekhin
and S.V. Urazhdin, J. Phys.: Condens. Matter \textbf{10}, 9561
(1998).
\bibitem{Takahasi}H. Takahasi, J. Phys. Soc. Jap. \textbf{16}, 1685 (1961).

\bibitem{Turik}A.V. Turik, Ferroelectrics \textbf{221}, 111 (1999).


\end{thebibliography}
\end{document}